\documentclass[aps,superscriptaddress,prd,aps,
reprint,
floatfix,
amsmath,amssymb,amsfonts]{revtex4-2}
\usepackage{amsmath,amssymb}
\usepackage{mathrsfs}
\usepackage[dvipdfmx]{graphicx}
\usepackage{tikz}
\usepackage{bm}
\usepackage{subfigure}
\usepackage{here}
\usepackage{subfigure}
\usepackage{float}
\usepackage{comment}
\usepackage{dcolumn}
\usepackage{bm,float}
\usepackage[legacycolonsymbols]{mathtools}

\newcommand\nn{\nonumber \\}
\newcommand\dd[1]{d{#1}\ }
\newcommand{\pa}{\partial}

\DeclareMathOperator{\Ei}{Ei}

\newcommand{\anno}[1]{\textcolor{red}{#1}}
\newcommand{\annob}[1]{\textcolor{blue}{#1}}
\usepackage{ulem}
\usepackage{xcolor}
\DeclareRobustCommand{\erase}{\bgroup\markoverwith{\textcolor{red}{\rule[.5ex]{2pt}{0.8pt}}}\ULon}
\usepackage{cancel}

\usepackage[colorlinks,linkcolor=blue,anchorcolor=violet,citecolor=red]{hyperref}

\begin{document}
\title{Detector Based Evaluation of Extractable Entanglement in Flat spacetime}

\author{Hiromasa Tajima}
\email{tajima.hiromasa.m6@s.mail.nagoya-u.ac.jp}
\affiliation{ Department of Physics, Nagoya University, Nagoya 464-8602, Japan}
\author{Riku Yoshimoto}
\email{yoshimoto.riku.d1@s.mail.nagoya-u.ac.jp}
\affiliation{ Department of Physics, Nagoya University, Nagoya 464-8602, Japan}
\author{Ryo Nemoto}
\email{nemotoryo@eken.phys.nagoya-u.ac.jp}
\affiliation{ Department of Physics, Nagoya University, Nagoya 464-8602, Japan}
\author{Yuki Osawa}
\email{osawa.yuki.e8@s.mail.nagoya-u.ac.jp}
\affiliation{ Department of Physics, Nagoya University, Nagoya 464-8602, Japan}
\begin{abstract}
  Entanglement entropy (EE) is widely used to quantify quantum correlations in field theory, with the well-known result in two-dimensional conformal field theory (CFT) predicting a logarithmic divergence with the ultraviolet (UV) cutoff. However, this expression lacks operational meaning: it remains unclear how much of the entanglement is physically extractable via local measurements. In this work, we investigate the operationally accessible entanglement by employing a pair of Unruh-DeWitt detectors, each interacting with complementary regions of a quantum field. We derive an upper bound on the entanglement that can be harvested by such detectors and show that it scales as a double logarithm with respect to the UV cutoff—significantly weaker than the single-logarithmic divergence of the standard CFT result. This work provides an operational perspective on field-theoretic entanglement and sets fundamental limits on its extractability.
\end{abstract}
\maketitle

\section{Introduction and Summary}
Quantum systems are known to exhibit nonlocal correlations known as quantum entanglement, which cannot be explained by classical physics and can exceed the limits of classical correlations, as demonstrated by violations of Bell inequalities.
One useful quantity to quantify quantum entanglement among quantum systems $AB$ is the entanglement entropy $S_{EE}$ defined as $S_{EE}:=-\mathrm{Tr}[\hat{\rho}_A\log{\hat{\rho}_A}]$ where $\hat{\rho}_A$ is the reduced density matrix for the quantum system $A$.
When the total system $AB$ is pure, this entanglement entropy has clear operational meaning: it corresponds to the asymptotic number of EPR-like Bell pairs that can be extracted from many copies of the system using local operations and classical communication (LOCC).
In such extraction protocols, ancillary systems—referred to as probes—that interact locally with $A$ or $B$, may be introduced to facilitate the transfer and storage of quantum information.

In the 2-dimensional conformal field theory (CFT), the EE of the ground state (vacuum state) for a single interval of length is $L$ and its complementary is famously given by 
\begin{align}
    S_{\mathrm{EE}}
    =\frac{c}{3}\log\biggl(\frac{L}{\epsilon}\biggr),
    \label{eq:EE-replica}
\end{align}
where $c$ is the central charge of CFT and $\epsilon$ is the ultraviolet (UV) regulator
\cite{srednicki1993entropy,Holzhey:1994we,calabrese2004entanglement}, which can be derived by replica method
(A useful review is here \cite{Nishioka:2018khk,Kusuki:2024gtq}). 
As $\varepsilon\to0$, the entanglement entropy diverges logarithmically, indicating that the local degrees of freedom inside the interval share an infinite amount of entanglement with their complement. 
This divergence raises important questions regarding the operational meaning of EE:
Can all of this entanglement be extracted in a realistic physical process? If not, how much of it is operationally accessible?

To address these questions, we must clarify the notions of probes and local operations within the framework of the quantum field theory.
In this context, an Unruh-DeWitt (UDW) detector \cite{Unruh:1976db,dewitt1979,Birrell:1982ix} serves as probes, and the local operation on the quantum field is defined by its interaction with the detector. 
By employing two or more such detectors, each locally interacting with the field and spatially separated from one another, one can swap entanglement from the quantum field---shared among different spacetime regions---into the detectors. This protocol is known as entanglement harvesting\cite{reznik2003entanglement,reznik2005violating,kukita2017harvesting, deSLTorres:2023ujd}, and has been investigated in the relativistic quantum information (RQI) community.

In this work, we investigate the maximum of entanglement which can be extracted using a pair of UDW detectors, where one detector interacts with the quantum field within an interval of length $L$, and the other interacts with its compliment. 
As we shall discuss later, UDW detector can be characterized by a pair of smeared field operators, referred to as detector modes or the local modes\cite{Hotta:2015yla,trevison2019spatially,yamaguchi2020superadditivity, yamaguchi2020superadditivity,nambu2023analog,Osawa:2024fqb,osawa2025entanglement}. 
Therefore, instead of specifying the full internal structure of the detector, one can define it operationally by choosing appropriate smearing (or window) functions for the detector modes. 
For a detector designed to measure within a given interval, its smearing function is taken to have support only inside that region. Conveniently, such functions can be expanded in terms of Fourier series.
We examine the possible Fourier coefficients of these smearing function and demonstrate that the entanglement extractable by a pair of such detectors is bounded by a ~$\log\log$-law~\eqref{eq:EE}, whose UV divergence is weaker than the EE of the interval \eqref{eq:EE-replica}.
 
The rest of the paper consists as follows. We explain some concepts of the detector model in Sec.~\eqref{sec:Detector_mode}.
Subsequently, in Sec.~\eqref{sec:evedetetorEE}, we evaluate the upper bound of the EE which can be extracted by using pair of UDW detectors in the interval and its complementary, which is the main result of this paper.
Sec.~\eqref{sec:Conclusion} is devoted to summary and future perspective.

\section{Detector Mode}
\label{sec:Detector_mode}
In this section, we briefly review detector-based measurements and local operations of the quantum field. For a more thorough review and additional details, see~\cite{Osawa:2024fqb,osawa2025entanglement}.
First, we introduce the concept of particles and particle detectors or UDW detectors.
Then, we introduce detector modes, which represent local operations on the field.

In globally hyperbolic spacetime, the vacuum state of the field and the concept of particles are defined only after we specify the constant‑time (Cauchy) slice \cite{Birrell:1982ix,wald1994quantum}.
In the standard framework of quantum field theory in Minkowski spacetime, where only inertial observers are considered, such ambiguity does not arise, thanks to the invariance of the vacuum under Poincar\'{e} transformations connecting different inertial observers \cite{weinberg1995quantum}. 
However, this is not the case when non-inertial observers or gravity are involved. To address this ambiguity, Unruh and DeWitt proposed defining particles as excitations of the UDW detectors \cite{Unruh:1976db,dewitt1979}. Their definition generalizes the particle concept in the standard Minkowski quantum field theory. 
It is important to note that this particle concept depends not only on the choice of the Cauchy slice but also on the details of the UDW detectors (e.g., structure of the coupling, switching function, etc.).
Moreover, in Minkowski spacetime, the detector-based particle mode can be related to the inertial particle mode through a Bogoliubov transformation.

UDW detectors or the local operation on the quantum field by using them can generally be characterized by the interaction Hamiltonian \cite{trevison2019spatially,yamaguchi2020superadditivity,nambu2023analog}:
\begin{align}
  H_{\mathrm{int}}(\tau)
  = \lambda \,\eta(\tau) \, f(\hat{q}(\tau),\hat{p}(\tau)) \hat{O}_D(\tau),
  \label{eq:H_int}
\end{align}
where $\tau$ is the proper time of the detector, 
$\hat{q}$ and $\hat{p}$ are smeared field operator, and $\hat{O}_D(\tau)$ is a detector observable. The time dependence of the operators indicates Heisenberg picture.
Also, $\lambda$ and $\eta(\tau)$ denote the coupling constant and 
the switching function of the interaction, respectively, and $f({q},{p})$ is a real function of $q$ and $p$.
The smeared field operators determine the degrees of freedom within the quantum field whose time evolution is affected by the interaction defined by a given $\hat{H}_{\text{int}}$. They can be chosen arbitrarily, reflecting the properties of the system, as long as they satisfy the canonical commutation relation $\bigl[\hat{q},\hat{p}\bigr]=i$. 

Since the operators $\hat{q}$ and $\hat{p}$ are chosen to satisfy canonical commutation relation, they are regarded as canonical variables of the smeared field.
The meaning of these variables becomes clearer by introducing creation and annihilation operators defined as
\begin{align}
    \hat{A}
    \coloneqq \frac{\hat{q}+i\hat{p}}{\sqrt{2}},
\end{align}
with the corresponding creation operator given by the Hermitian conjugate to $\hat{A}$.
These operators $\hat{A},\hat{A}^\dagger$ satisfy the standard commutation relation for creation and annihilation operators:
\begin{align}
\bigl[\hat{A},\hat{A}^\dagger\bigr]=1.
\end{align}
It is worth noting that by employing the mode expansion of the field operator:
\begin{align}
    \hat{\phi}(\tau,\xi)=\int\dd{\omega} \left\{\hat{a}_\omega\varphi_\omega(\tau,\xi)+h.c.\right\},
\end{align}
the operator $\hat{A}$ and the set $\left\{\hat{a}_\omega\right\}_\omega$ is related via the Bogoliubov transformation.
In this sense, the pair of canonical variables $\{\hat{q},\hat{p}\}$, which is determined by the details of the interaction, defines a mode or equivalently a notion of particles. The mode defined through this procedure is referred as detector mode.

Indeed, all information about the quantum field that can be extracted by the detector is encapsulated in the detector mode or its smearing functions. Therefore, the specific details of the detector such as $\hat{O}_D^I$ or energy level are no longer necessary, once the detector mode is specified.
Thus, in the remainder of this paper, we define a local probe simply by specifying the smearing functions of the detector mode, rather than detailing the internal structure of the detector.

\section{Entanglement Entropy for the detector mode}\label{sec:evedetetorEE}
In this section, we evaluate the amount of entanglement extractable by using a pair of UDW detectors, one interacting with the quantum field in a finite interval and the other with its complementary region.
As UDW detectors effectively swap entanglement with the corresponding detector modes, the entanglement between these detector modes corresponds to entanglement extractable via the UDW detectors.
Furthermore, from the perspective of the partner formula\cite{Hotta:2015yla,trevison2019spatially}, this entanglement is maximized when one detector mode coincides with the partner mode of the other. 
Therefore, in the following, we evaluate the entanglement entropy between a detector mode in a finite interval  and its partner mode, which provides the upper bound on the amount of extractable entanglement in this setup.

In this work, we focus on massless chiral scalar field (the field with only right-moving or left-moving degrees of freedom) in 1+1 dimensional Minkowski spacetime, for simplicity. 
For such chiral fields, it is natural to choose a constant $v=t+x$ null surface as a Cauchy surface, since the field evolves solely along the $u=t-x$ direction.
In this case, smeared field operators are generally given by \cite{tomitsuka2020partner,Osawa:2024fqb,osawa2025entanglement}
\begin{align}
    \hat{q}\coloneqq\int\dd{u} Q(u)\hat{\Pi}(u),\quad
    \hat{p}\coloneqq\int\dd{u} P(u)\hat{\Pi}(u),
    \label{eq:smeared_canonical_variable}
\end{align}
where $\hat{\Pi}(u):=\pa_u\hat{\phi}(u)$ is the canonical momentum conjugate to $\hat{\phi}(u)$.
The smearing functions $Q(u)$ and $P(u)$ are chosen  to satisfy $\int \dd{u}Q(u)P'(u)=-2$, which ensures canonical commutation relation.

Please note that a finite spatial interval $x\in[-L/2, L/2]$ at $t=0$ corresponds directly to a finite interval $u\in [-L/2, L/2]$ on the $v=0$ null slice.
This correspondence allows us to describe localized measurements or operations over spatial intervals using smeared operators defined along null coordinates.

Generally, the detector mode whose window function is non-zero in $u\in [-L/2,L/2]$ can be represented by using Fourier series:  
\begin{align}
\begin{split}
    &\hat{q}
    = \int_{-L/2}^{L/2}\dd{u}
    \sum_{n\ne0}
      Q_ne^{-2in\pi u/L}\hat{\Pi}(u)\\
    &\hat{p}
    = \int_{-L/2}^{L/2} \dd{u}
    \sum_{n\ne0}P_ne^{-2in\pi u/L}\hat{\Pi}(u),
\end{split}
\end{align}
where, we have omitted $n=0$ component. This does not affect generality, as it corresponds to a constant shift in the window function, which does not change the canonical commutation relation.
The Fourier component $Q_n$ and $P_n$ is constrained to satisfy the canonical commutation relation of $\hat{q}$ and $\hat{p}$,
\begin{equation}
    i=[\hat{q},\hat{p}]=i\sum_{n\ne 0}Q_nP_n^*(i n \pi),
\end{equation}
and the reality condition for the smearing  functions 
\begin{align}
    Q_{-n}=Q_n^*, P_{-n}=P_n^*
\end{align}
are also imposed for the Fourier components.

The annihilation operator is defined by
\begin{align}  \label{eq:ani}
  \hat{A}
  &=\int_{-L/2}^{L/2} \dd{u}
    \sum_{n\ne0}
    \frac{Q_n+iP_n}{\sqrt{2}}
    e^{-2in\pi u/L}\hat{\Pi}(u)\nn
  &=-i\sum_{n\ne 0}(-1)^nC_n\nn
  &\,\times\int_0^{\infty}
  \frac{\dd{\omega}}{\sqrt{4\pi\omega}}\left[\frac{2 \sin(L\omega/2)}{1+2n\pi/L\omega}\hat{a}_{\omega}-\frac{2 \sin(L\omega/2)}{1-2n\pi/L\omega}\hat{a}_{\omega}^{\dagger}\right]\nn
        &\eqqcolon\int_{0}^{\infty}\dd{\omega}\left(\alpha_{\omega}(\omega)\hat{a}_{\omega}+\beta_{\omega}(\omega)\hat{a}_{\omega}^{\dag}\right).
\end{align}
    We defined $C_n=(Q_n+i P_n)/2$ and \eqref{eq:ani} can be transformed into the following form
    \begin{align}
        \label{eq:standardform}
        \hat{A}=\alpha\hat{a}_{\tilde{\parallel}}+\gamma\hat{a}_{\tilde{\parallel}}^\dag+\delta\hat{a}_{\perp}^\dag.
    \end{align}
   where $\hat{a}_{\tilde{\parallel}}$ is the annihilation part of $\hat{A}$, and other parameters are defined by
\footnotesize
\begin{align}
\begin{split}
  \hat{a}_{\tilde{\parallel}}
  &=\frac{1}{\alpha}
    \int \dd{\omega}
    \alpha_{\omega}(\omega)
    \hat{a}_{\omega},\quad   \alpha=\sqrt{\int\dd{\omega}|\alpha_{\omega}(\omega)|^2},\\
        \gamma\ &=[\hat{a}_{\tilde{\parallel}},\hat{a}_A]=\frac{1}{\alpha}\int\dd{\omega}\alpha_{\omega}(\omega)\beta_{\omega}(\omega),\\ \delta \ &=\sqrt{\alpha^2-|\gamma|^2-1},\quad
        \hat{a}_{\perp}=\frac{1}{\delta}\left(\int \dd{\omega}\beta_{\omega}^{*}(\omega)\hat{a}_{\omega}-\gamma^{*}\hat{a}_{\tilde{\parallel}}\right)
\end{split}
\end{align}
\normalsize
To eliminate the single-mode squeezing in $\hat{A}$, we perform the local symplectic transformation 
\begin{equation}\label{eq:symplectic trans}
\hat{a}_{\parallel}
= \cosh{\tilde{r}}\hat{a}_{\tilde{\parallel}}
- e^{i\phi}\sinh{\tilde{r}}
  \hat{a}_{\tilde{\parallel}}^{\dagger}.
\end{equation}
where $\tilde{r}$ and $\phi$ are given by
 $\tanh{\tilde{r}}
  =-{|\gamma|}/{\alpha}$ and $\phi =\text{Arg} \ \gamma$.
With this transformation \eqref{eq:symplectic trans}, we obtain 
\begin{align}
  \hat{A}
  = \cosh{r}\hat{a}_{\parallel}
  + \sinh{r}\hat{a}_{\perp}^{\dagger}
\end{align}
where the parameter $r$ is given by
\begin{equation}
  \tanh{r}
  = \frac{\delta
    }{\alpha \cosh{\tilde{r}}
    +|\gamma|\sinh{\tilde{r}}}.
\end{equation}

The entanglement entropy between the detector mode and its complementary is determined by the squeezing parameter of the detector mode \cite{trevison2018pure,trevison2019spatially}:
\begin{align}
    \label{eq:EElocal}
  S_{EE}
  = \sqrt{1+g^2}
    \log
    \left(
     \frac{\sqrt{1+g^2}+1}{|g|}
    \right)
  + \log\left(\frac{|g|}{2}\right)
\end{align}
where $g=\sinh 2r$. To evaluate $S_{\mathrm{EE}}$, we have to calculate the integrals in $\alpha$ and $\gamma$ 
\footnotesize
\begin{align}
\begin{split}
  & \hspace{-3mm}
  \int_0^{\infty}\dd{\omega}
  \bigl|\alpha_{\omega}(\omega)\bigr|^2\\
  &= \sum_{\substack{n,n'\ne0}}
    \int_0^{\infty}\frac{d\omega}{\pi}
    \frac{
      (-1)^{n+n'}C_nC_{n'}^*\times
    \omega e^{-\epsilon \omega}\sin^2{{L \omega}/{2}}
    }{(\omega+{2n\pi}/{L})
      (\omega+{2n'\pi}/{L})
    },\end{split}\\
  \begin{split}
    & \hspace{-3mm} 
    \int_0^{\infty}\dd{\omega}
    \alpha_{\omega}(\omega)\beta_{\omega}(\omega)\\
  &=
   \sum_{\substack{n,n'\ne0}}
   \int_0^{\infty}\frac{d\omega}{\pi}
    \frac{
      (-1)^{n+n'}C_nC_{n'}^*\times\omega e^{-\epsilon \omega}\sin^2{{L \omega}/{2}}
    }{
      (\omega+{2n\pi}/{L})(\omega+{2n'\pi}/{L})
    }.
    \end{split}
\end{align}
\normalsize
In this calculation, we introduced the cutoff parameter $\epsilon$ to regularize the UV divergence. Using the calculations in the Appendix, the asymptotic behavior of $\alpha^2$ and $|\gamma|^2$ in $L/\epsilon\to\infty$ are given by
\footnotesize
\begin{align}\label{eq:alpha}
  \hspace{-3.4mm}
  \alpha^2&\approx\hspace{-.3mm}\frac{1}{2\pi}
    \left(\sum_{\substack{n,n'\ne 0}}
    (-1)^{n+n'}
    C_nC_{n'}^*\right)
  \log\frac{L}{\epsilon},\\
  \hspace{-3.4mm}
  \label{eq:gamma}
  |\gamma|^2&
  \approx\hspace{-.3mm}
    \frac{1}{2\pi}\left(\frac{
      \left|
      \displaystyle{\sum_{{n,n'\ne 0}}}
        (-1)^{n-n'}C_nC_{-n'}
      \right|^2
    }{
         \displaystyle{\sum_{\substack{n,n'\ne 0}}}
      (-1)^{n+n'}C_nC_{n'}^*}\right)
  \log\frac{L}{\epsilon}.
\end{align}
\normalsize

Substituting \eqref{eq:alpha} and \eqref{eq:gamma} into \eqref{eq:EElocal}, we obtain the asymptotic behavior of the entanglement entropy:
\begin{align}
\label{eq:EE}
    S_{EE}
    \approx 
    \log\left(\log \frac{L}{\varepsilon}\right).
\end{align}
The EE of the detector modes in the 2-dimensional flat spacetime exhibits a weaker UV divergence behavior compared to the EE of the finite region of the 2-dimensional CFT. 
This indicates that the amount of entanglement extracted by using a pair of UDW detectors is significantly smaller than the total entanglement present in the quantum field restricted to a finite interval. 

\section{Conclusion and outlooks}
\label{sec:Conclusion}
In this Letter, we evaluated the amount of entanglement which can be extracted by using a pair of detectors, with one interacting with quantum field in a finite interval and the other interacting with its complementary region.
We showed that the extractable entanglement follows double logarithmic dependence ~\eqref{eq:EE} on the UV regulator, which exhibits weaker UV divergence compared to the well-known logarithmic behavior ~\eqref{eq:EE-replica} of entanglement entropy in conformal field theory.
Our result indicates that it is not possible to extract the full vacuum entanglement by using a single pair of UDW detectors.

In our setup, we considered a single pair of UDW detectors. However, a finite interval of size $L$ contains approximately $L/\varepsilon$ independent modes, where $\varepsilon$ is the UV cutoff.
Hence, we expect that the logarithmic law of entaglement entropy may be recovered if multiple pairs of UDW detector are employed.
However, when generalizing in this way, we must recall that our present analysis likely overestimates the amount of extractable entanglement.
A naive estimate of the total entanglement—given by (number of independent modes)$\times$(EE per mode)$\sim L/\varepsilon\times\log\log L/\varepsilon$—would exceed the standard logarithmic law.
This overestimation arises because multiple detector modes within an interval generally share quantum entanglement, and such internal or “self-entanglement” must be subtracted when evaluating the total entanglement between the interval and its complement.
While properly accounting for this self-entanglement is nontrivial, it presents an interesting direction for future investigation.

Our result~\eqref{eq:EE}, at first glance, seems apparently different from the Ryu-Takayanagi(RT) formula~\cite{Ryu:2006bv}.
The RT formula says that the EE in the finite region of the CFT can be interpreted as the area entropy of the RT surface in its bulk region, 
and the EE calculated by the RT formula means the geometry of the bulk region surrounded by the RT surface~\cite{Dong:2016eik}. 
This interpretation of the RT surface is consistent with our result that the field has the bound entanglement, that cannot be extracted by the detector.
Therefore, the geometrical meaning for our result from the holographic aspects is of interest to us.
Moreover, it is also interesting to examine whether \eqref{eq:EE} reproduces key features of EE—such as strong subadditivity, which can be proven geometrically in the RT formula.

We should keep it mind that in our settings we only consider the EE in flat spacetime and do not treat the gravitational effect. Generally, in the curved spacetime where the gravitational effect is turned on, particle creation occurs, such as Hawking radiation in the black hole spacetime. The smearing function is also influenced by the gravity because the finite region of spacetime is curved by the gravitational effect.
Gravity affects some parts of our setup, so the result can be non-trivial. 

To deal with this uncertainty by gravity, it may be insightful to study our settings in the context of algebraic QFT. For instance, it was shown by \cite{Chandrasekaran:2022cip} that in the de Sitter spacetime, types of von Neumann factor of QFT in static patch can be changed from type III$_{1}$ to type II$_{1}$ by taking the observer's degrees of freedom into account. Mathematically, this operation is identified as the crossed product, and physically it can be interpreted as adding gravity. Results in \cite{Chandrasekaran:2022cip} imply that after introducing an observer in the de Sitter spacetime, we are able to define the trace and the EE can take finite values. Moreover, it turned out that the EE matches generalized entropy $S_{\text{gen}}=(A/4G_{N})+S_{\text{out}}$. Thus, it is interesting to calculate the EE with and without an observer on the de Sitter spacetime by means of our settings, and see their differences from the information theoretical perspectives. 

In this discussion, we should pay attention to some conditions. The most important thing is that there are some assumptions for observers in \cite{Chandrasekaran:2022cip} such that its mass is infinite and they are localized along geodesics on north or south poles. Under this assumption, it is not clear that how our calculations should be changed. Recently, however, some investigations have been done by loosening some assumptions, such as \cite{kolchmeyer2024chaos}. It may be possible to reconfigure out calculations with such works in mind.

These insights will be useful to interpret~\eqref{eq:EE} from various points of view.
By using them, we are expecting to reach deep understandings of the operational meaning of the EE.

\section*{Acknowledgments}
We would like to thank Yasusada Nambu and Masanori Tomonaga for their useful comments. This work was financially supported by JST SPRING, Grant Number JPMJSP2125. H.T, R.N. and Y.O. would like to take this opportunity to thank the ``THERS Make New Standard Program for the Next Generation Researchers.''

\appendix
\section{The detail of the calculation}
We summarize the integrals used in the main text. 
\footnotesize
\begin{widetext}
\begin{align}
    &\int_0^{\infty}\dd{\omega}
  \frac{\omega\sin^2{{L\omega}/{2}}}{(\omega+{2 n\pi}/{L})^2}= \frac{e^{{2n\pi \epsilon}/{L}}}{2}\int_{2n\pi}^{\infty}\dd{\omega}\left(\frac{1-\cos{\omega}}{\omega}-2n\pi\frac{1-\cos{\omega}}{\omega^2}\right)e^{-{\epsilon\omega}/{L}}\\
    &\int_0^{\infty}\dd{\omega}\frac{\omega\sin^2{{L\omega}/{2}}}{(\omega+{2 n\pi}/{L})(\omega+{2 n'\pi}/{L})}
    =\frac{ne^{{2n\pi \epsilon}/{L}}}{2(n-n')}\int_{2n\pi}^{\infty}\dd{\omega}\frac{1-\cos{\omega}}{\omega}e^{-{\epsilon\omega}/{L}}
    -\frac{n'e^{{2n'\pi \epsilon}/{L}}}{2(n-n')}\int_{2n'\pi}^{\infty}\dd{\omega}\frac{1-\cos{\omega}}{\omega}e^{-{\epsilon\omega}/{L}}\\
   &\int_{2n\pi}^{\infty}\dd{\omega}\frac{1-\cos{\omega}}{\omega}e^{-{\epsilon\omega}/{L}}
   =-\Ei\left[-\frac{2n\pi\epsilon}{L}\right]
  -\frac{1}{2}\left\{\Ei\left[-2n\pi\left(-i+\frac{\epsilon}{L}\right)\right]+\Ei\left[-2n\pi\left(i+\frac{\epsilon}{L}\right)\right]\right\}\\
   &\int_{2n\pi}^{\infty}\dd{\omega}\frac{1-\cos{\omega}}{\omega^2}e^{-{\epsilon\omega}/{L}}=-\frac{\epsilon}{L}\Ei\left[-\frac{2n\pi\epsilon}{L}\right]
   -\frac{1}{2}\left\{\left(-i+\frac{\epsilon}{L}\right)\Ei\left[-2n\pi\left(-i+\frac{\epsilon}{L}\right)\right]\right.
   \left. 
   +\left(i+\frac{\epsilon}{L}\right)\Ei\left[-2n\pi\left(i+\frac{\epsilon}{L}\right)\right]\right\}
\end{align}
\end{widetext}
\normalsize
 where $\Ei(x)$ is the exponential integral function defined as \vspace{-3mm}
\footnotesize
\begin{align}
    \Ei(x):=P\int_{-\infty}^x\dd y\frac{e^y}{y}=-P\int^{\infty}_x\dd y\frac{e^{-y}}{y}.
\end{align}
\normalsize
Here, $P$ denotes the operation taking principal value of the integral, however we can ignore it unless $x=0$. The asymptotic behavior of $\Ei(x)$ around $x\sim0$ is
\footnotesize
\begin{align}
    \Ei(x)\approx \log |x|+\gamma+x \quad \gamma\text{ : Euler's }\gamma\text{ constant}.
\end{align}
\normalsize
Adopting this asymptotic behavior of $\mathrm{Ei}(x)$, we obtain the leading term of the entanglement entropy \eqref{eq:EE}
    
\bibliographystyle{unsrt}
\bibliography{ref.bib}

\end{document}